\documentclass[12pt]{article}
\usepackage[T1]{fontenc}

\usepackage{amssymb,amsmath}
\usepackage{graphics}
\usepackage{subfig, graphicx, caption}
\graphicspath{{images/}}
\usepackage{color}
\usepackage{array,multirow}
\usepackage{epsfig}
\usepackage{mwe}
\usepackage{comment}
\usepackage[a4paper]{geometry}
\usepackage{cite}
\usepackage{placeins}
\usepackage[utf8]{inputenc}
\usepackage{cancel}
\usepackage{arydshln}
\makeatletter \renewcommand{\@dotsep}{10000} \makeatother
\usepackage{indentfirst}
\usepackage[colorlinks=true
,urlcolor=blue
,citecolor=blue
,linkcolor=blue
,pagecolor=blue
,linktocpage=true
,pdfproducer=medialab
]{hyperref}
\def\beq{\begin{equation}}
\def\eeq{\end{equation}}
\usepackage[nodisplayskipstretch]{setspace}

\begin{document}

\begin{titlepage}

\begin{flushright}

\end{flushright}
\pagestyle{empty}

\begin{center}
{\Large \bf The anomalous $Wtb$ vertex and top-pair production at the LHC}\\
\vspace{8pt}
{\bf  Apurba Tiwari\footnote{\tt E-mail: atiwari@myamu.ac.in} and Sudhir Kumar Gupta\footnote{\tt E-mail:sudhir.ph@amu.ac.in} }
\vspace{2pt}
\begin{flushleft}
{\em Department of Physics, Aligarh Muslim University, Aligarh, UP--202002, INDIA} 
\end{flushleft}

\vspace{10pt}
\begin{abstract}
We perform an exclusive study of the anomalous $Wtb$ interaction in the context of LHC. The limits on $Wtb$ anomalous couplings have been 
estimated via the measurements of the top-quark decay width and cross-section as well as production asymmetries in case the CP is violated. 
Our investigations reveal that the upper bounds on $(|C_L|, |C_R|)$ would be of about (1.82, 0.03) $\times 10^{-4}$ at 
$2.5 \sigma$ C.L., for the already collected data at the LHC with $\sqrt{S} = 13$ TeV for an integrated luminosity of 139 fb$^{-1}$. The 
corresponding limits for future hadron colliders, namely, High Luminosity LHC (HL-LHC), High Energy LHC (HE-LHC) and Future Circular 
Collider (FCC-hh) for the projected luminosities of 3 ab$^{-1}$, 12 ab$^{-1}$ and 30 ab$^{-1}$ are found to be to 
$(0.81, 0.006)\times 10^{-4}$, $(0.44, 0.0017)\times 10^{-4}$ and $(0.21, 0.0004)\times 10^{-4}$ respectively at 2.5$\sigma$ C. L..
\end{abstract}
\end{center}

\end{titlepage}

\section{Introduction} 
\label{intro} 
The Standard-Model (SM) \cite{Novaes:1999yn} is so far considered to be a highly successful theory to interpret most of the data collected in 
experiments. Nevertheless, SM is still an incomplete model due to the fact that it remains infelicitous in explaining phenomenon such as 
$CP$-violation \cite{Grossman:1997pa,Stecker:2002tb}, Leptogenesis \cite{Davidson:2008bu}, Baryogenesis \cite{Cline:2006ts} and the existence 
of dark matter and dark energy \cite{Sahni:2004ai}. Besides it also lacks gravitational interaction \cite{Minkevich:2008jdt}, does not 
provide an adequate solution to the hierarchy problem \cite{Jegerlehner:2013nna} etc which hints for extending the SM to explain some or all 
of the problems listed above. The importance of the $CP$-violation searches lies in the fact that it will help to understand the 
matter-antimatter asymmetry present in the Universe. Also, the observation of $CP$-violation would be a clear indication of physics 
Beyond-the-SM (BSM) because the observed amount of $CP$-violation in SM is very tiny which is not sufficient to understand the current 
matter-antimatter asymmetry \cite{Gronau:1996wg}.

The $CP$-violating interactions involving the top-quark are quite promising as the top-quark could be a direct probe to such phenomenon partly 
due to its shorter life-time ($10^{-25}$ sec) and much heavier mass than the other quarks which enables it to decay much before 
hadronisation alike other quarks. For this reason, the final decay products are expected to preserve information about the top-quark 
properties and hence precise study of the final decay products would be useful and informative for direct physics searches. Also, top-quark 
sector is one of the most promising areas for precision studies as it provides direct probes to the interactions to the BSM physics. Precision 
studies from the past have predominated to explore new physics contributions to kinematically inaccessible regions. With the introduction of 
the Large Hadron Collider (LHC) \cite{Maestre:2021ajm}, precision measurements in top physics are directed towards a new era. Since LHC is now 
producing an enormous number of top quarks per year and is known as the "top-quark factory", it is considered a prominent place for the 
study of top-quark properties.

In the present article, we explore the $CP$ properties of the anomalous $Wtb$ vertex in the context of top-quark through the top-pair production process  and their consequent decays 
via the $W^\pm$ at the LHC and its proposed variants. Such interactions could be modelled by constructing the effective Lagrangian of higher dimension where the modification in the 
$Wtb$ vertex is through an extra term in the Lagrangian and where the $Wtb$ vertex is parameterised in terms of four unknown form factors \cite{Boos:2016zmp} which may in general be 
CP-even or CP-odd. As the CP-even contribution is expected to raise the event rates involving the aforementioned vertex both at decay as well as  production levels, sensitivities to such interactions could be estimated by employing the already measured values to these at the LHC. The CP-odd contribution is expected to interfere with the SM contribution and is expected to be observed in the form of asymmetries to the aforementioned processes which could be estimated constructing the CP-odd observables 
as suggested in the 
Refs.~\cite{Groote:2008ux,Gupta:2009wu,Han:2009ra,Gupta:2009eq,Hayreter:2013kba,Hayreter:2015cia,Hayreter:2015ryk,Tiwari:2019kly}. The contribution to anomalous couplings via 
top-quark production and decay has already been extensively studied in the existing 
literature and can be found in Refs. \cite{ATLAS:2012nhi,CMS:2013xxb,CMS:2014uod,ATLAS:2015ryj,Rindani:2015vya,Aguilar-Saavedra:2006qvv,
RomeroAguilar:2015plc,Dutta:2013mva,Alioli:2017ces,Jueid:2018wnj,MohammadiNajafabadi:2008llf,Aguilar-Saavedra:2008quj,D0:2009jke,Cao:2015doa,Cirigliano:2016njn,Cirigliano:2016nyn}. 
The $Wtb$ vertex structure has been studied in multiple ways in previous studies, for example: the $Wtb$ vertex has been studied using the 
$W$-boson helicity fractions \cite{ATLAS:2012nhi,CMS:2013xxb,CMS:2014uod}, through the $t$-channel cross-section \cite{ATLAS:2015ryj}, via 
the polarisation of top-quark \cite{Rindani:2015vya}, using angular asymmetries and helicity fractions individually 
\cite{Aguilar-Saavedra:2006qvv}. The $CP$-violating observables sensitive to $Wtb$ vertex have been discussed in 
Ref. \cite{RomeroAguilar:2015plc}. In Ref. \cite{Alioli:2017ces}, the anomalous $Wtb$ vertex has been explored in the context 
of collider, flavour observables and low energy measurements as well as electric dipole moments and in Ref. \cite{Dutta:2013mva}, the 
CP-conserving anomalous $Wtb$ couplings have been studied through single top production at $e^{-}p$ collider. It has been noticed in earlier studies 
\cite{Birman:2016jhg,Bernardo:2014vha,Deliot:2017byp} that combining measurements obtained from different studies 
can place comparatively stringent limits on anomalous couplings than those obtained from individual measurements. The LHC bounds on such 
anomalous interactions have also been measured by both CMS and ATLAS by means of top-decay asymmetry and single top production 
\cite{Aguilar-Saavedra:2011jfc}.

Organisation of the paper is as follows: In Section \ref{process_lag} we present the Lagrangian containing the anomalous $Wtb$ interactions 
and its consequences for the top-quark decay as well as top-pair production process at the LHC. Section \ref{Num_Anal} discusses constraints 
on the anomalous $Wtb$ interactions through the measurements of top-quark decay width and its pair production cross-section at the LHC. We 
also construct CP-asymmetries involving the $t\bar{t}$ production process at the LHC and its proposed variants and finally estimate the 
sensitivities corresponding to each of these. Finally, we summarise our findings in Section \ref{summary}.

\section{Process and Lagrangian construction} 
\label{process_lag} 

As mentioned earlier, the focus of present study is to explore the $CP$-violating interactions of the top-quark with $W$-boson and $b$-quark 
at the LHC and other hadron colliders. Since the SM does not allow anomalous interactions, such effects are incorporated in an effective 
field theoretic manner by extending the SM $Wtb$-vertex through the following effective Lagrangian \cite{delAguila:2002nf},  
\begin{eqnarray}
\mathcal L_{Wtb} = -\frac{g}{\sqrt {2}} \bar{b} \left[\gamma^{\mu}\left(C_{1L}P_L + C_{1R} P_R\right)W^{-}_{\mu} 
- i\sigma^{\mu\nu}\left(\tilde C_{2L} P_L + \tilde C_{2R} P_R\right) (\partial_{\nu} W^{-}_{\mu})\right]t + h.c.
\label{eff_lag}
\end{eqnarray}
where, $\tilde C_{2L, 2R} = \frac{C_{2L, 2R}}{\Lambda}, P_{L, R} = \frac{1}{2} (1 \mp \gamma_5)$, and $\Lambda$ is the energy scale. 
$C_{1L}, C_{1R}, C_{2L}$ and $C_{2R}$ are dimensionless complex anomalous couplings known as form factors. For the SM $Wtb$ vertex, value of $C_{1L}$ is equal to $V_{tb}$ and can be taken as unity with the assumption of 
unitarity of the CKM matrix, and the other anomalous couplings are equal to zero. Obviously non-zero values of either of the aforementioned 
couplings would indicate the presence of BSM interaction of the top-quark. However, we focus only on the anomalous couplings, namely, 
$C_{2L}$ and $C_{2R}$ and hence the rest of the couplings (other than $C_{1L}$) are set to zero. Also for the sake of simplicity from here 
onwards, we will call $C_{2L, 2R}$ as $C_{L, R}$. It is to be noted that in the limit of b-quark being massless, a left-handed W-boson contributes to the top-decay and for the decay of an anti-top quark, a left-handed W-boson is 
forbidden and a right-handed W-boson contributes. Therefore top-quark decay proceeds through left-handed charged current interactions and 
an anti-top quark proceeds through right-handed charged current interaction. Also, the analysis with $C_L \neq 0, C_R = 0$ and 
$C_R \neq 0, C_L = 0$ represent left-handed and right-handed currents respectively.

In order to explore such interactions at the LHC, it is clearly worthwhile to consider processes involving top-pair production, as LHC is 
designated to be a top factory wherein about $28.8$ M top-quarks have been pair produced for a Centre-of-mass (CMS) energy of $\sqrt{S}$ = 13 TeV 
at the LHC so far. It is to be further noted that among the produced tops about $87\%$ are through the fusion of gluon pairs while the rest 
are due to the annihilation of a pair of quark and anti-quark. As the top-quark (which is heaviest among all the quarks) has a life span 
which is much shorter than the time required for it to get hadronised, unlike other quarks, it does not form any bound states and hence it is 
expected to serve as a direct probe to $CP$ violation or any other anomalous effects. This, therefore, yields us ample hope to investigate such 
effects at colliders through the top-quark allied processes. As the anomalous $Wtb$ vertex has a tensorial character, probing such 
interactions requires reconstruction of the full partonic process relevant for the given LHC process, however since it is not always 
feasible partially due to the missing neutrinos. Besides, the complexity further rises to another level due to the involvement of information 
about the spin and polarisation of each of the particles participating in a given process. However, it has been recently shown that by 
deploying T-odd triple products involving only momenta of the final state particles of a given process, one could still explore such 
anomalous interactions \cite{Groote:2008ux,Gupta:2009wu,Han:2009ra,Gupta:2009eq,Hayreter:2013kba,Hayreter:2015cia,Hayreter:2015ryk,Tiwari:2019kly} at hadron colliders. However, in the present study in order to explore the maximum experimental sensitivities 
in the context of the LHC and other hadron colliders, we will only utilise the absolute asymmetries.

\section{Numerical Analysis} 
\label{Num_Anal}

As the anomalous $Wtb$ vertex is expected to affect the production as well as decay processes, we first consider decay of top-quark and 
later investigate the production process $p p \longrightarrow t (\to b W^+) {\bar t} (\to {\bar b} W^-)$ at the LHC.

For our calculations, we incorporated the Lagrangian mentioned in Section \ref{process_lag} into the 
${\tt FeynRules}$ \cite{Christensen:2008py} and later the model files generated by ${\tt FeynRules}$ were interfaced to 
${\tt FeynCalc}$ \cite{Mertig:1990an} for further investigations. As the anomalous couplings $C_{L, R}$ are complex, we may have two types of 
contributions due to the anomalous interactions, namely the CP-even or CP-odd with former affecting the rates while the latter reflecting in 
terms of the asymmetries to the processes discussed above. In either case, in order to ensure that the rates are in good agreement with 
their respective measurements, we first demand that the contribution due to such interactions should lie within the experimental 
uncertainties. 
	
The decay level CP asymmetries could be estimated by  
\begin{eqnarray}
{\cal A}^{\Gamma}_{SM} &=& \frac{\Delta \Gamma_{t\to b W}}{\Gamma_{t\to b W}} \simeq \frac{{\rm Im} \left(|{\cal M}|^2_{t\to b W}\right) } 
{{\rm Re} \left(|{\cal M}|^2_{t\to b W}\right)} 
\label{as_gamx}
\end{eqnarray}
with $ |{\cal M}|^2_{t\to b W}$ being the matrix-element squared for the process $t\to b W$. From the above expression it is obvious to note
that ${\cal A}^{\Gamma}_{SM} = 0$ due to the absence of anomalous $Wtb$ coupling within the SM.

Before proceeding further, we have already verified the following result for the relative decay width of the top-quark in presence of the 
anomalous couplings to the SM decay width as mentioned in Ref. \cite{Agrawal:2012ga}, 

\begin{eqnarray}
R^{\Gamma} &=& \frac{\Gamma_{t\to b W}}{\Gamma^{SM}_{t\to b W}} = 1 - \frac{M_W}{(1 + 2 \eta^2)}[6 \eta C_R - M_W(\eta^2 + 2)(C_L^2 + C_R^2)]
\label{Rgam}
\end{eqnarray}
where $\eta = \frac{M_W}{m_t}$, $C_L = |C_L| e^{i\theta}$ and $C_R = |C_R| e^{i\phi}$.
\begin{figure}[]
\centering
\includegraphics[width=0.45\linewidth]{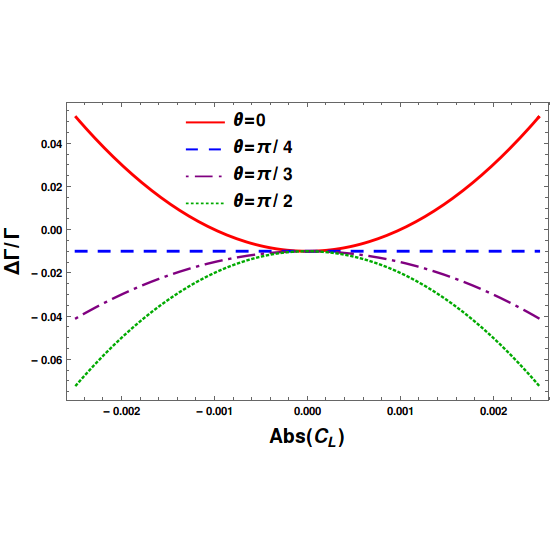}
\includegraphics[width=0.45\linewidth]{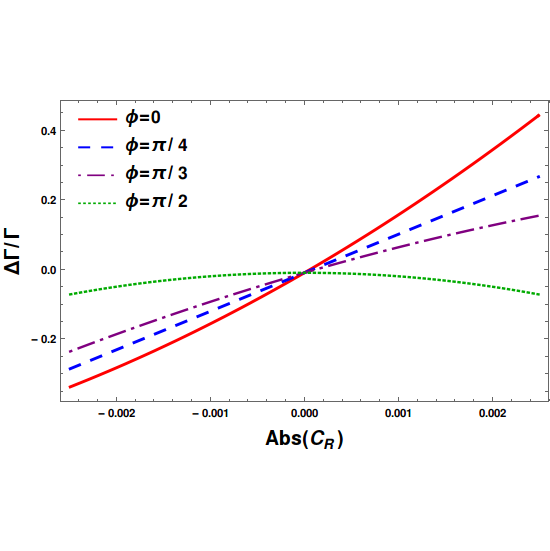}
\caption{\small Decay width of top-quark as a function of moduli of anomlaous couplings for different values of phases $\theta$ and $\phi$  
for the cases with  $|C_R|$ = 0 (left) and $|C_L|$ = 0 (right) where $C_L = |C_L| e^{i\theta}$ and $C_R = |C_R| e^{i\phi}$.}
\label{decywidth_thetdependency}
\end{figure}

In Fig. \ref{decywidth_thetdependency}, we show the relative change in decay width as a function of moduli of the anomalous coupling at
different values of phases $\theta$ and $\phi$, for illustration we consider $\theta = \phi = 0, \frac{\pi}{4},
\frac{\pi}{3}$ and $\frac{\pi}{2}$, for the cases with $|C_R|$ = 0 (left) and $|C_L|$ = 0 (right). The decay-width 
does not seem to show any appreciable change by varying Abs$(C_L)$. Furthermore, the curve is almost symmetric around
Abs$(C_L)$ = 0 which means that the decay width is equally sensitive to both positive and negative values of the moduli. The symmetric curve
indicates that the contribution to the linear order is very tiny. This is because no linear term corresponding to the coupling
$C_L$ exists in the relative decay width expression \ref{Rgam} and the contribution is only from the quadratic term $C_L^{2}$. Conversely,
the decay width changes significantly on varying the values of Abs$(C_R)$, indicating the strong dependence of the decay width on the
anomalous coupling $C_R$. However, the sensitivity of the decay width to Abs$(C_R)$ apparently depends on the choice of values of $\phi$, for 
example, the decay width is more sensitive to positive values of Abs$(C_R)$ for $\phi$ = 0 and becomes almost equally sensitive to both 
positive and negative values of Abs$(C_R)$ for $\phi$ = $\frac{\pi}{4}$ and $\frac{\pi}{3}$. Although, the variation in Abs$(C_R)$ has 
almost no impact on the decay width at $\phi$ = $\frac{\pi}{2}$, whereas the decay width can change by about $\pm 50\%$ on varying the values 
of Abs$(C_R)$ at $\phi$ = $\frac{\pi}{4}$ and $\frac{\pi}{3}$. The strong dependence of the decay width on the coupling $C_R$ is factually 
due to the presence of the term proportional to $C_R$ in the relative decay width expression \ref{Rgam}. Clearly, by examining these results, 
a rough estimate of the constraints on the phase $\phi$ and on Abs$(C_R)$ can be presented to obtain the most significant value of the 
top-quark decay width.

The limits on anomalous couplings $C_L$ and $C_R$ at 2.5$\sigma$ confidence level (C.L.) obtained from the top-quark decay width measurements 
are given in Table \ref{cnstrnt_dcyxsec}, where we assume only one anomalous coupling to be non-zero at a time.  

\begin{table}[]
\centering
\scalebox{0.8}{
\renewcommand{\arraystretch}{2.0}
\begin{tabular} {c|c|c}
\hline
                                                                        & $C_L~(\times 10^{-3})$     & $C_R~(\times 10^{-3})$ \\
\hline\hline
$\left(\frac{\Delta \Gamma}{\Gamma}\right)_{t \to bW}$                  & $ -5.86 \le C_L \le 5.86 $ & $ -1.84 \le C_R \le 1.95 $\\
$\left(\frac{\Delta \sigma}{\sigma}\right)_{pp \to t\bar{t}}^{13 TeV}$  & $ -2.62 \le C_L \le 2.62 $ & $ -0.40 \le C_R \le 0.40 $\\
\hline\hline
\end{tabular}}
\caption{Individual constraints on anomalous couplings $C_L$ and $C_R$ (when only one anomalous coupling is taken non-zero at a time) at 
2.5$\sigma$ C.L. obtained from measurements of top-quark decay width and top-pair production cross-section at the LHC with  
$\sqrt{S}$ = 13 TeV.}
\label{cnstrnt_dcyxsec}
\end{table}

\begin{figure}[]
\centering
\includegraphics[width=0.45\linewidth]{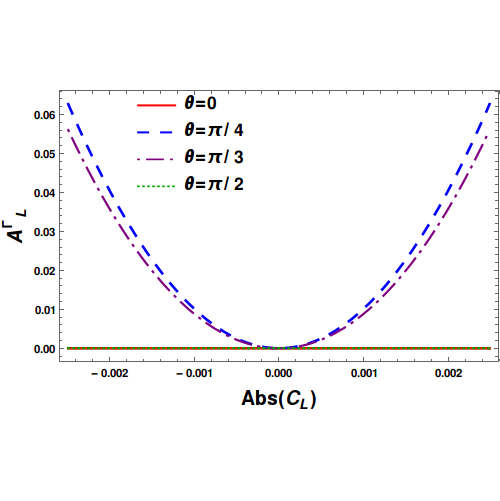}
\includegraphics[width=0.45\linewidth]{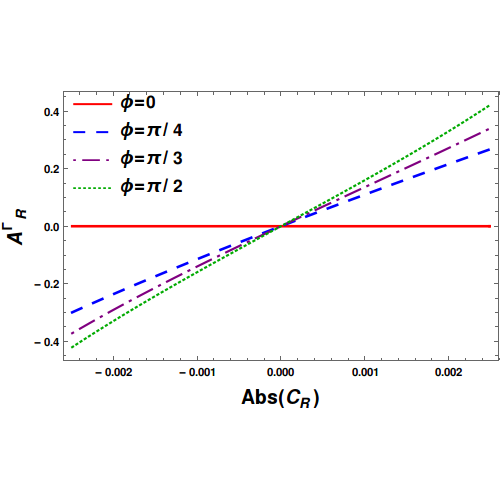}
\caption{\small Dependence of decay level asymmetry on the moduli of anomalous couplings $C_L$ and $C_R$ for the cases with
$|C_R|$ =0 (left) and $|C_L|$ =0 (right).}
\label{asdecywidth_thetdpndcy}
\end{figure}
In Fig. \ref{asdecywidth_thetdpndcy}, we show decay level asymmetry ($A^{\Gamma}$) as a function of moduli of the anomalous coupling for 
the cases with $|C_R|$ =0 (left) and $|C_L|$ =0 (right). Four illustrative values of phases, $\theta$ = $\phi$ = 0, $\frac{\pi}{4}$, 
$\frac{\pi}{3}$ and $\frac{\pi}{2}$ are considered. We see that asymmetry is almost symmetric around Abs($C_L$) = 0 and Abs($C_R$) = 0, 
signifying that it is equally sensitive to the positive and negative values of the moduli of the couplings $C_L$ and $C_R$. 
Also, we find that the contribution from coupling $C_L$ is insignificant and that the asymmetry is more sensitive to 
coupling $C_R$. Furthermore, for larger values of $\phi$, the lines become steeper which means that larger values of 
$\phi$ $(\sim \frac{\pi}{2})$ make a more significant contribution.

\begin{figure}[]
\centering
\includegraphics[width=0.45\linewidth]{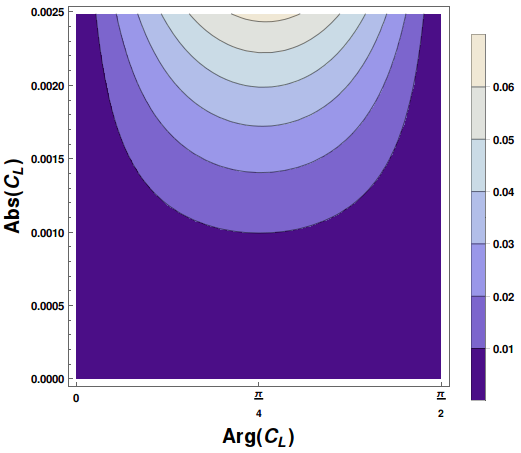}
\includegraphics[width=0.45\linewidth]{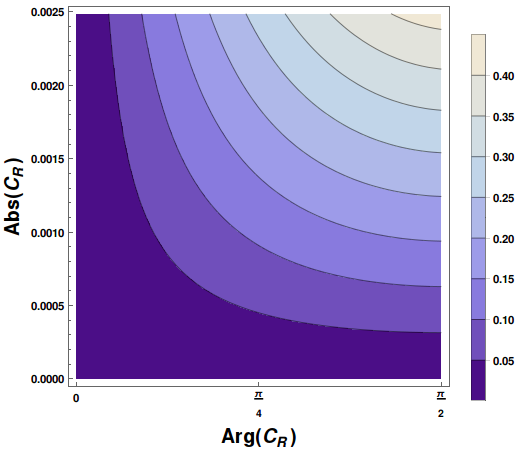}
\caption{\small Contour plots of decay level asymmetry in the Abs($C_L$)-Arg($C_L$) plane and Abs($C_R$)-Arg($C_R$) plane for the
cases with $|C_R|$ =0 (left) and $|C_L|$ =0 (right).}
\label{cntr_plt_asdecywidth}
\end{figure}

In Fig. \ref{cntr_plt_asdecywidth}, we show the contour plots of decay level asymmetry in the Abs($C_L$)-Arg($C_L$) plane and Abs($C_R$)-Arg($C_R$)
plane for the cases with $|C_R|$ =0 (left) and $|C_L|$ =0 (right). Here also, we see that coupling
$C_L$ makes a negligible contribution to the asymmetry whereas $C_R$ shows a meaningful contribution to it, again
indicating that asymmetry is more sensitive to coupling $C_R$. 

\begin{figure}[]
\centering
\includegraphics[width=0.45\linewidth]{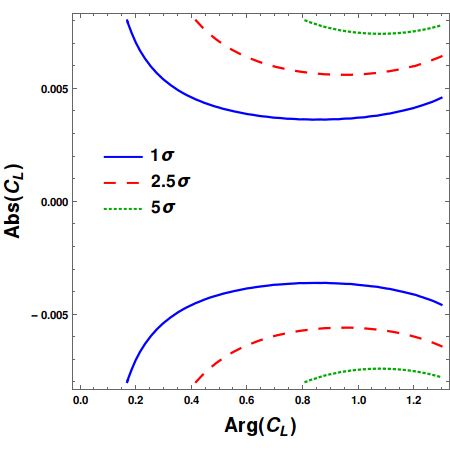}
\includegraphics[width=0.45\linewidth]{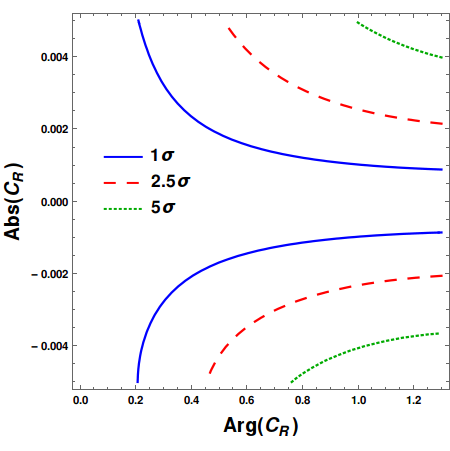}
\caption{\small The 1$\sigma$, 2.5$\sigma$ and 5$\sigma$ C.L. regions in the Abs($C_L$)-Arg($C_L$) plane and Abs($C_R$)-Arg($C_R$) plane
allowed by the decay level asymmetry for the cases with $|C_R|$ =0 (left) and $|C_L|$ =0 (right).}
\label{cntr_siglvl_decywidth}
\end{figure}

In Fig. \ref{cntr_siglvl_decywidth}, we show the 1$\sigma$, 2.5$\sigma$ and 5$\sigma$ regions in Abs($C_L$)-Arg($C_L$) plane and
Abs($C_R$)-Arg($C_R$) plane allowed by the decay level asymmetry for the cases with $|C_R|$ =0 (left) and $|C_L|$ =0 (right). From Fig. \ref{cntr_siglvl_decywidth}, we 
can give a rough estimate of the limits on the moduli and phase of the anomalous couplings at 2.5$\sigma$ C.L.. However, it does not make a significant contribution as the asymmetry
obtained at the decay level is very small.

\begin{figure}[]
\centering
\includegraphics[width=0.45\linewidth]{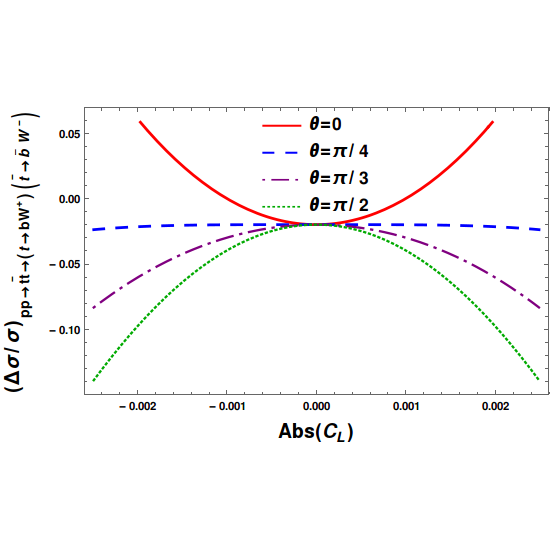}
\includegraphics[width=0.45\linewidth]{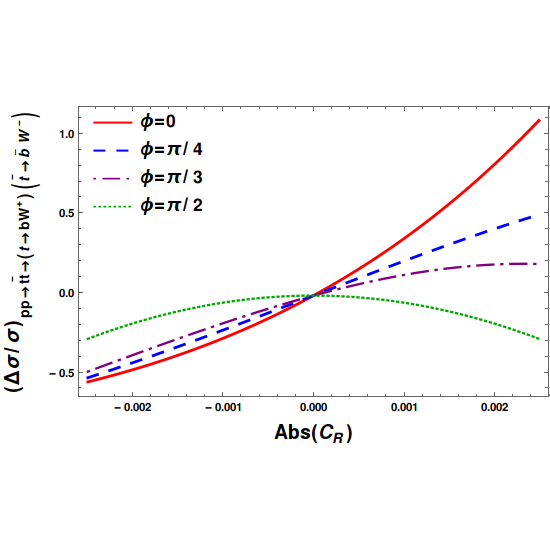}
\caption{\small Cross-section as a function of moduli of anomalous couplings for different values of phases $\theta$ and $\phi$ for the 
cases with $|C_R|$ =0 (left) and $|C_L|$ =0 (right) where $C_L = |C_L| e^{i\theta}$ and $C_R = |C_R| e^{i\phi}$.}
\label{xsec_thetdepndncy}
\end{figure}

In order to obtain the constraints on $C_R$ and $C_L$ at the cross-section level, we first estimate the LHC cross-sections for the process 
$ p p\longrightarrow t (\to b W^+) {\bar t} (\to  {\bar b} W^-)$ using {\tt MadGraph5} \cite{Alwall:2011uj} after implementing the 
contribution due to the already existing $tbW$ interaction in the {\tt FeynRules}. We then compare this with the SM cross-section which is 
generated by switching-off $C_L$ and $C_R$. The additional contribution to the SM cross-section $\Delta \sigma = \sigma - \sigma_{SM}$ is then obtained by keeping $C_L$ or $C_R$ non-zero at a time as shown in Fig. \ref{xsec_thetdepndncy}. The individual constraints on anomalous couplings $C_L$ and $C_R$ at 2.5$\sigma$ C.L. are then obtained from the measurement of top-pair production cross-section at the LHC at 
$\sqrt{S}$ = 13 TeV. These are presented in Table \ref{cnstrnt_dcyxsec}, where we assume only one anomalous coupling to be non-zero at a 
time. 

The Fig. \ref{xsec_thetdepndncy} shows the relative change in cross-section, $\frac{\Delta{\sigma}}{\sigma}$ as a function of moduli of the 
anomalous coupling at four different values of $\theta$ and $\phi$, namely $\theta$ = $\phi$ = $0, \frac{\pi}{4}, 
\frac{\pi}{3}~\rm{and}~\frac{\pi}{2}$ for the cases with $|C_R|$ =0 (left) and $|C_L|$ =0 (right). We notice that 
just as for the decay width, the cross-section is more sensitive to the values of coupling $C_R$. We see that changing the values of the 
coupling $C_L$ has a small effect on the cross-section but is equally sensitive to both positive and negative values of coupling $C_L$. 
However, in the case of $C_R$, the cross-section dependence varies with the value of the phase of anomalous coupling, $\phi$. For example, 
for $\phi$ = 0 the cross-section is more sensitive to positive values of $C_R$, for $\phi$ = $\frac{\pi}{4}$ and $\frac{\pi}{2}$ it is 
equally sensitive to both positive and negative values of the coupling $C_R$ and for $\phi$ = $\frac{\pi}{3}$ it is more sensitive to negative values 
of $C_R$. From Fig. \ref{xsec_thetdepndncy}, we see that the change in cross-section is highest when Abs$(C_R)$ is changed for $\phi$ = 0, 
indicating that the correct choice of the value of $\phi$ can lead to a significant change in the cross-section. 

\begin{figure}[]
\centering
\includegraphics[width=0.45\linewidth]{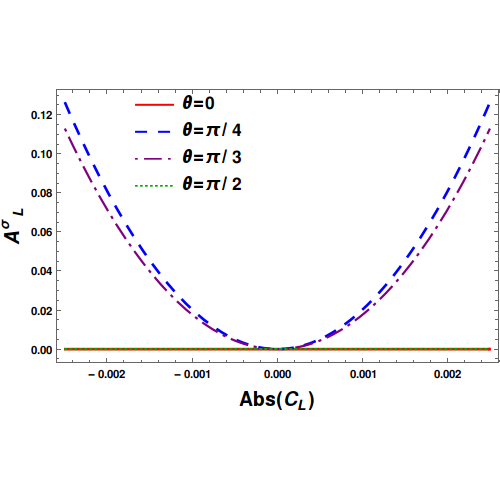}
\includegraphics[width=0.45\linewidth]{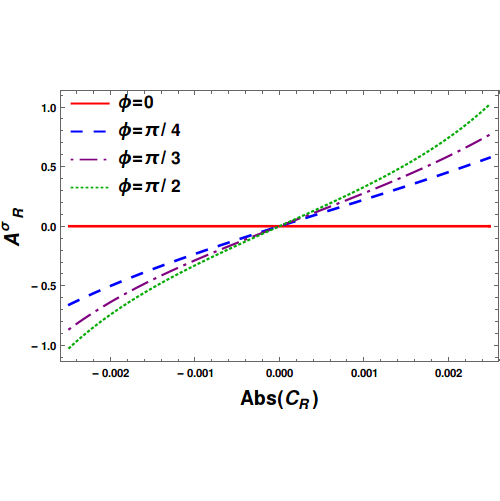}
\caption{\small Dependence of production asymmetry on the moduli of anomalous couplings for different values of $\theta$ and $\phi$ for the 
cases with $|C_R|$ = 0 (left) and $|C_L|$ = 0 (right).}
\label{asxsec_thetdpndcy}
\end{figure}

\begin{figure}[]
\centering
\includegraphics[width=0.45\linewidth]{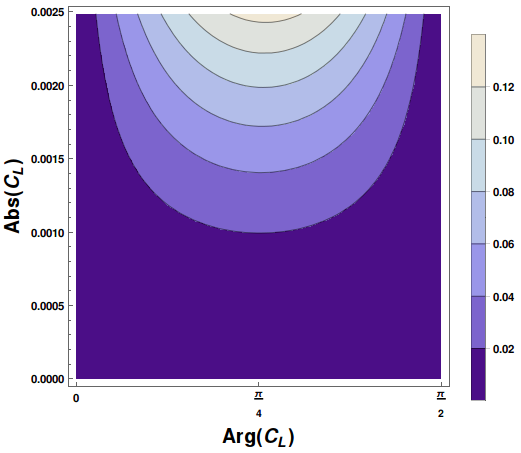}
\includegraphics[width=0.45\linewidth]{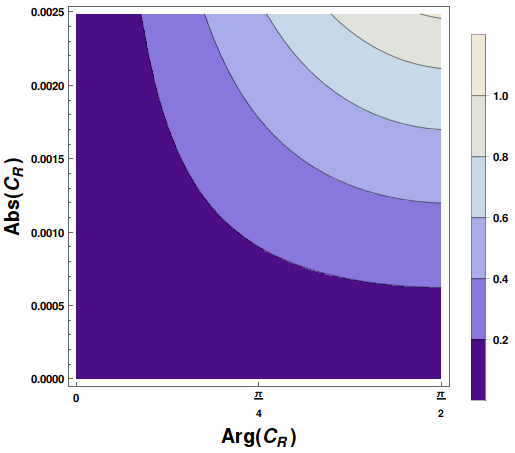}
\caption{\small Contour plots of production asymmetry in the Abs($C_L$)-Arg($C_L$) plane and Abs($C_R$)-Arg($C_R$) plane for the 
cases with $|C_R|$ =0 (left) and $|C_L|$ =0 (right).}
\label{cntr_plt_asxsec}
\end{figure}
We now turn our discussion for the LHC sensitivities corresponding to the couplings $C_L$ and $C_R$. The production asymmetries could be 
estimated in the same way as the decay level asymmetries using
\begin{eqnarray}
{\cal A}^{\sigma}_{SM} &=& \frac{\Delta \sigma_{p p \longrightarrow t (\to b W^+) {\bar t} (\to {\bar b} W^-)}}
{\sigma_{p p \longrightarrow t (\to b W^+) {\bar t} (\to {\bar b} W^-)}} \simeq \left(\frac{{\rm Im} \left(|{\cal M}|^2_{t\to b W}\right)}
{{\rm Re} \left(|{\cal M}|^2_{t\to b W}\right)}\right)^2.
\label{as_sigx}
\end{eqnarray}
In Fig. \ref{asxsec_thetdpndcy}, we show the dependence of production asymmetry ($A^{\sigma}$) on moduli of the anomalous couplings $C_L$ and 
$C_R$ for the cases with $|C_R|$ = 0 (left) and $|C_L|$ = 0 (right). To illustrate the impact of CP-violating phases on production asymmetry, 
we consider four different values of phases, $\theta = \phi = 0,~\frac{\pi}{4},~\frac{\pi}{3}~\rm{and}~\frac{\pi}{2}$. From 
Fig. \ref{asxsec_thetdpndcy}, we notice that the asymmetry is almost insensitive to coupling $C_L$ and significantly sensitive to coupling 
$C_R$. Since the coupling $C_L$ does not contribute much to the asymmetry, we will not go into its detailed explanation and focus on the 
effect of the CP-violating phases of the coupling $C_R$ on the production asymmetry. From Fig. \ref{asxsec_thetdpndcy} (right), we see that 
the asymmetry is zero for $\phi$ = 0, increases significantly on increasing the value of the phase and reaches its maximum value for 
$\phi~=~\frac{\pi}{2}$. Also, we find that the asymmetry is equally sensitive to both positive and negative values of the moduli of the 
anomalous coupling.

Fig. \ref{cntr_plt_asxsec} shows the contour plots of production asymmetry in the Abs($C_L$)-Arg($C_L$) plane and Abs($C_R$)-Arg($C_R$) 
plane for the cases with $|C_R|$ = 0 (left) and $|C_L|$ = 0 (right). The plots again justify that the
asymmetry is more sensitive to coupling $C_R$ and that the contribution from coupling $C_L$ is much smaller. We observe that the maximum 
contribution to production asymmetry is at a value of $\phi~=~\frac{\pi}{2}$, which further confirms the results shown in 
Fig. \ref{asxsec_thetdpndcy}. Analysing Figs. \ref{asxsec_thetdpndcy} and \ref{cntr_plt_asxsec}, we conclude that the coupling $C_R$ is 
dominating over $C_L$ and the major contribution to the production asymmetry is from the coupling $C_R$.

\begin{figure}[]
\centering
\includegraphics[width=0.44\linewidth]{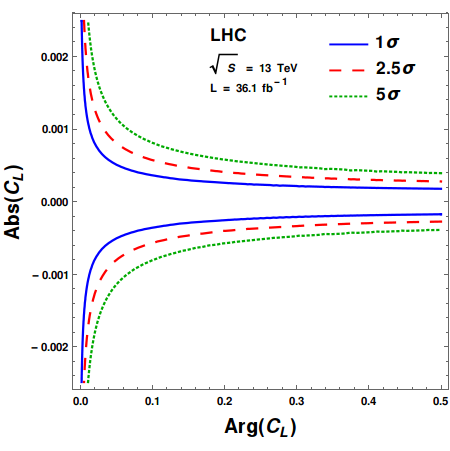}
\includegraphics[width=0.44\linewidth]{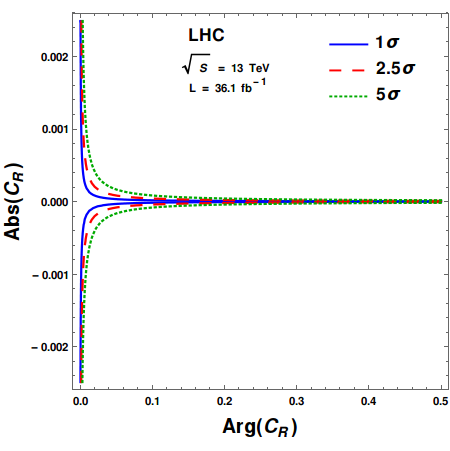}
\includegraphics[width=0.44\linewidth]{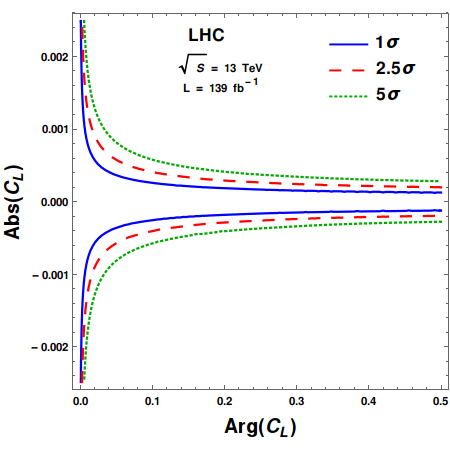}
\includegraphics[width=0.44\linewidth]{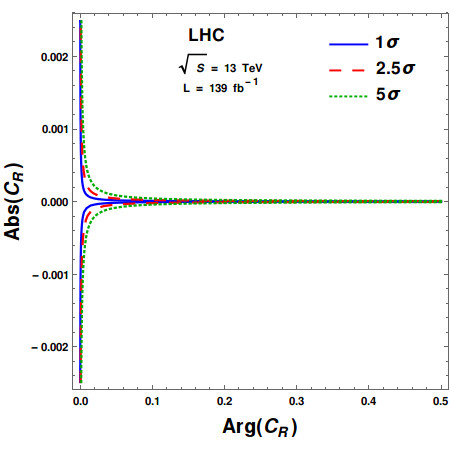}
\caption{\small 1$\sigma$, 2.5$\sigma$ and 5$\sigma$ C.L. regions in the Abs($C_L$)-Arg($C_L$) plane and Abs($C_R$)-Arg($C_R$) plane allowed
by the production asymmetry at CMS energy of $\sqrt{S}$ = 13 TeV for the integrated luminosities of 36.1 fb$^{-1}$ and 139 fb$^{-1}$ for the
cases with $|C_R|$ = 0 (left) and $|C_L|$ = 0 (right).}
\label{cntr_siglvl13TeV}
\end{figure}

\begin{figure}[]
\centering
\includegraphics[width=0.44\linewidth]{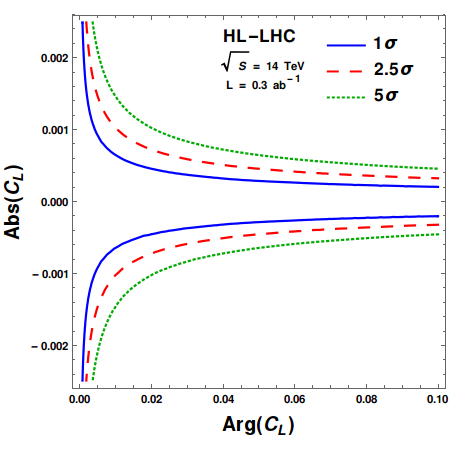}
\includegraphics[width=0.44\linewidth]{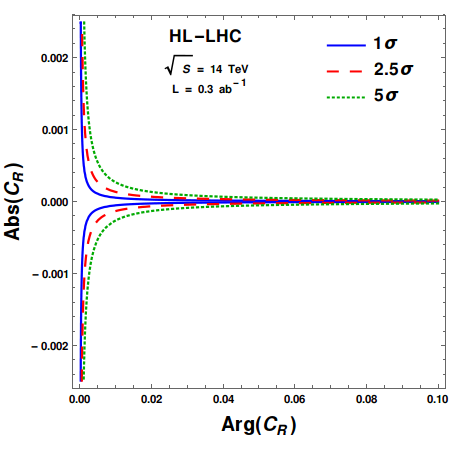}
\includegraphics[width=0.44\linewidth]{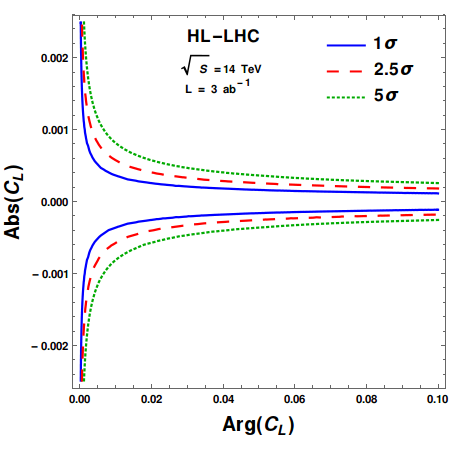}
\includegraphics[width=0.44\linewidth]{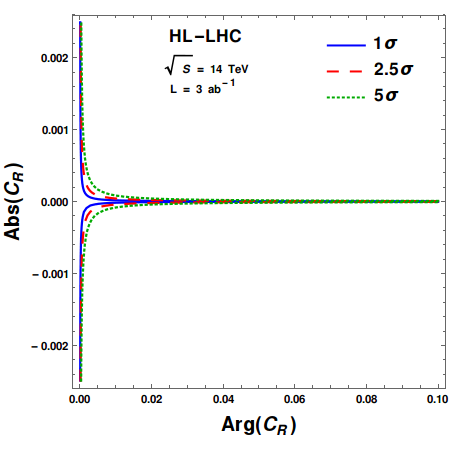}
\caption{\small 1$\sigma$, 2.5$\sigma$ and 5$\sigma$ C.L. regions in the Abs($C_L$)-Arg($C_L$) plane and Abs($C_R$)-Arg($C_R$) plane allowed
by the production asymmetry at CMS energy of $\sqrt{S}$ = 14 TeV for the integrated luminosities of 0.3 ab$^{-1}$ and 3 ab$^{-1}$
for the cases with $|C_R|$ = 0 (left) and $|C_L|$ = 0 (right).}
\label{cntr_siglvl14TeV}
\end{figure}

\begin{figure}[]
\centering
\includegraphics[width=0.44\linewidth]{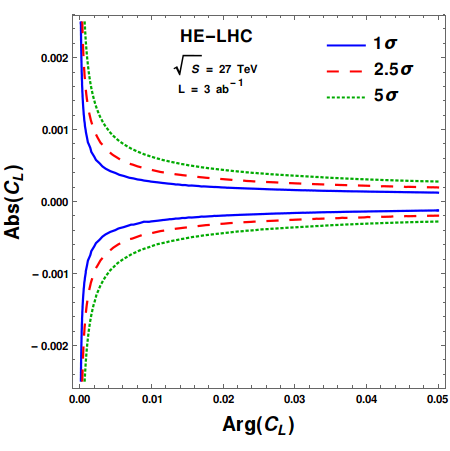}
\includegraphics[width=0.44\linewidth]{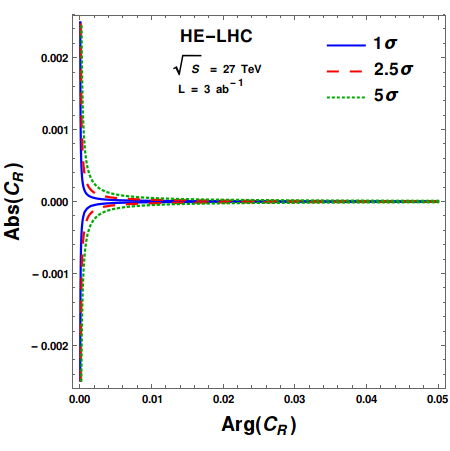}
\includegraphics[width=0.44\linewidth]{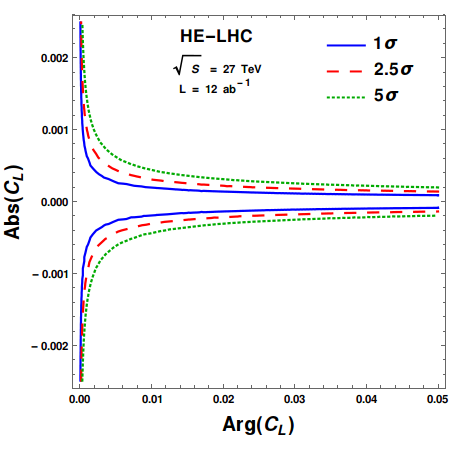}
\includegraphics[width=0.44\linewidth]{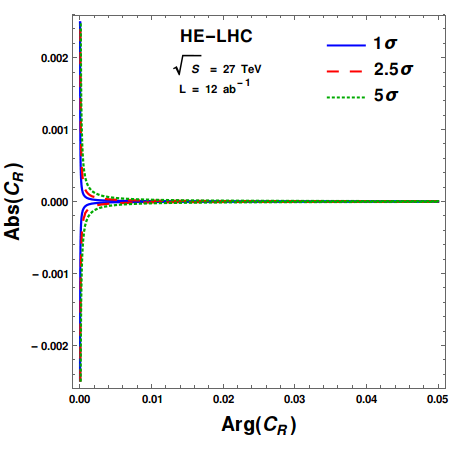}
\caption{\small 1$\sigma$, 2.5$\sigma$ and 5$\sigma$ C.L. regions in the Abs($C_L$)-Arg($C_L$) plane and Abs($C_R$)-Arg($C_R$) plane allowed
by the production asymmetry at CMS energy of $\sqrt{S}$ = 27 TeV for the integrated luminosities of 3 ab$^{-1}$ and 12 ab$^{-1}$ 
for the cases with $|C_R|$ = 0 (left) and $|C_L|$ = 0 (right).}
\label{cntr_siglvl27TeV}
\end{figure}

\begin{figure}[h]
\centering
\includegraphics[width=0.44\linewidth]{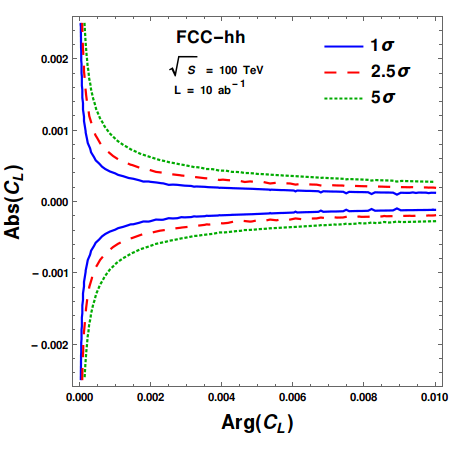}
\includegraphics[width=0.44\linewidth]{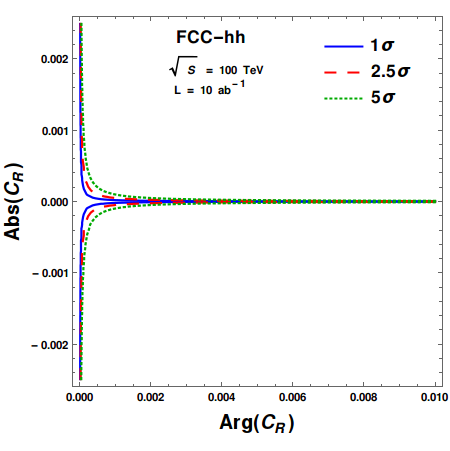}
\includegraphics[width=0.44\linewidth]{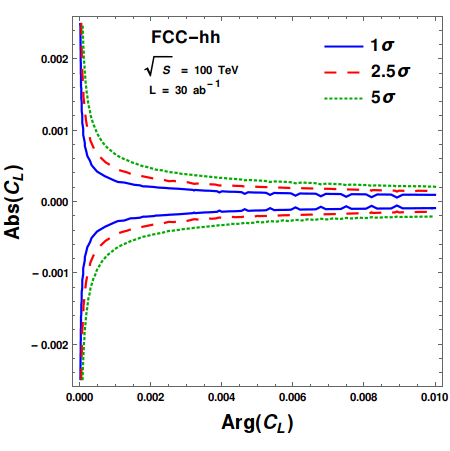}
\includegraphics[width=0.44\linewidth]{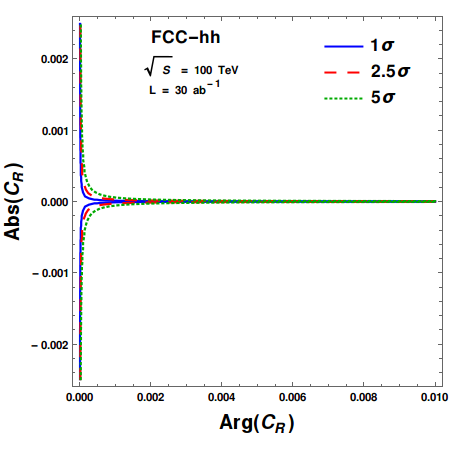}
\caption{\small 1$\sigma$, 2.5$\sigma$ and 5$\sigma$ C.L. regions in the Abs($C_L$)-Arg($C_L$) plane and Abs($C_R$)-Arg($C_R$) plane allowed
by the production asymmetry at CMS energy of $\sqrt{S}$ = 100 TeV for the integrated luminosities of 10 ab$^{-1}$ and 30 ab$^{-1}$ 
for the cases with $|C_R|$ = 0 (left) and $|C_L|$ = 0 (right).}
\label{cntr_siglvl100TeV}
\end{figure}

To estimate constraints on moduli and phase of the anomalous couplings, we first estimate the SM cross-sections using ${\tt MCFM-10.1}$ \cite{Campbell:2019dru} for various LHC energies as 
 required in our analysis. Later the sensitivities are estimated by comparing the obtained asymmetries against the error in the events through,

\begin{equation}
 A = \frac{n_{cl}}{\sqrt{\sigma_{p p \longrightarrow t (\to b W^+) {\bar t} (\to {\bar b} W^-)}^{SM} \times \int {\cal L} dt}}
\label{Asymm}
\end{equation}

where $\sigma_{p p \longrightarrow t (\to b W^+) {\bar t} (\to {\bar b} W^-)}^{SM}$ and $\int{\cal L} dt$ represent cross-section and 
integrated luminosity, respectively and $n_{cl}$ is the confidence level at which the bounds are to be obtained. For our investigations we 
use integrated luminosities of 36.1 fb$^{-1}$, 139 fb$^{-1}$ and 0.3 ab$^{-1}$, 3.0 ab$^{-1}$ for CMS energies of 13 TeV and 14 TeV, 
respectively. In addition, projected bounds on $C_{L, R}$ are also obtained for the proposed future hadron colliders, namely HE-LHC and 
FCC-hh with CMS energies of $\sqrt{S}$ = 27 TeV and 100 TeV with projected luminosities of 3.0 ab$^{-1}$, 12.0 ab$^{-1}$ and 10.0 ab$^{-1}$, 
30.0 ab$^{-1}$, respectively. The limits corresponding to each of these are shown in Table \ref{Cons_prodasymm}. In 
Figs. [\ref{cntr_siglvl13TeV}-\ref{cntr_siglvl100TeV}], we show the 1$\sigma$, 2.5$\sigma$ and 5$\sigma$ regions in Abs($C_L$)-Arg($C_L$) 
plane and Abs($C_R$)-Arg($C_R$) plane allowed by the production asymmetries for the cases with $|C_R|$ =0 (left) and $|C_L|$ =0 (right). We 
explore various energies at LHC, viz, $\sqrt{S}$ = 13 TeV, HL-LHC with $\sqrt{S}$ = 14 TeV, HE-LHC with $\sqrt{S}$ = 27 TeV and FCC-hh with 
$\sqrt{S}$ = 100 TeV. The luminosities of 36.1 fb$^{-1}$ to 30 ab$^{-1}$ have been explored. From 
Figs. [\ref{cntr_siglvl13TeV}-\ref{cntr_siglvl100TeV}], we can give a rough prediction of limits on the phase and moduli of the anomalous 
couplings at 2.5$\sigma$ C.L. However, the exact values of the limits have already been given in Table \ref{Cons_prodasymm}.

\begin{table}[h!]
  \begin{center}
\scalebox{0.8}{
  \renewcommand{\arraystretch}{1.6}
    \begin{tabular}{c|c|c|c}
Collider & $\sqrt{S}$, $\mathcal \int Ldt$ & $\left|C_L\right|~(\times 10^{-4})$ & $\left|C_R\right|~(\times 10^{-4})$ \\
\hline\hline
LHC      & 13 TeV, 36.1 fb$^{-1}$          &  2.55  & 0.06 \\
         & 13 TeV, 139 fb$^{-1}$           &  1.82  & 0.03 \\
\hline
HL-LHC   & 14 TeV, 0.3 ab$^{-1}$           &  1.44  & 0.019 \\
         & 14 TeV, 3.0 ab$^{-1}$           &  0.81  & 0.006 \\
\hline
HE-LHC   & 27 TeV, 3.0 ab$^{-1}$           &  0.62  & 0.0034 \\
         & 27 TeV, 12.0 ab$^{-1}$          &  0.44  & 0.0017 \\
\hline
FCC-hh   & 100 TeV, 10.0 ab$^{-1}$         &  0.28  & 0.0007 \\
         & 100 TeV, 30.0 ab$^{-1}$         &  0.21  & 0.0004 \\
\hline\hline
    \end{tabular}}
\caption{Individual bounds for CP-violating phase, $\theta=\phi=\frac{\pi}{4}$ on the moduli of the anomalous couplings $C_L$ and $C_R$ 
(when only one anomalous coupling is taken non-zero at a time) at 2.5$\sigma$ C.L. obtained from production asymmetries at LHC, HL-LHC, 
HE-LHC and FCC-hh.}
    \label{Cons_prodasymm}
  \end{center}
\end{table}

We now compare our results with the current limits on the anomalous couplings. In Ref. \cite{Birman:2016jhg}, the authors used W-boson 
helicities and single top-quark production cross-section measurements at LHC and Tevatron and the limits obtained on anomalous couplings at 
95$\%$ C.L. are: $-1.5\times 10^{-3} \le  \tilde g_R \le 1.1\times10^{-3}$, $-2.1\times 10^{-3} \le \tilde g_L \le 2.2\times10^{-3}$ at 
13 TeV LHC energy and $-0.87\times 10^{-3} \le \tilde g_R \le 0.87\times10^{-3}$, $-1.9\times 10^{-3} \le \tilde g_L \le 2.1\times10^{-3}$ 
at HL-LHC with $\sqrt{S}$ = 14 TeV. In Ref. \cite{Bernardo:2014vha}, the 95$\%$ C.L. limits obtained from W-boson helicities and t-channel 
cross-section at the LHC are: $-1.9\times 10^{-3} \le \tilde g_R \le 1.6\times10^{-3}$, 
$-1.4\times 10^{-3} \le \tilde g_L \le 1.0\times10^{-3}$, and from the combined measurements obtained at the LHC and Tevatron are: 
$-1.6\times 10^{-3} \le \tilde g_R \le 1.4\times10^{-3}$, $-1.2\times 10^{-3} \le \tilde g_L \le 0.9\times10^{-3}$. According to 
Ref. \cite{Deliot:2018jts}, the expected limits at 95$\%$ C.L. for HL-LHC for a CMS energy of 14 TeV and a total integrated luminosity of 
3 ab$^{-1}$ are: $-0.6\times 10^{-3} \le \tilde g_R \le 0.25\times10^{-3}$, $-2.1\times 10^{-3} \le \tilde g_L\le 2.4\times10^{-3}$ obtained 
from the combination of W boson helicity fractions, single top quark production cross sections and forward-backward asymmetries measured at 
Tevatron and at the LHC. 

\section{Summary and Conclusion} 
\label{summary}

We have discussed the effects of anomalous couplings in the $Wtb$ vertex on the decay of top-quark through its decay $t \to Wb$, and the 
production of $t{\bar t}$ pairs through their decays into $W$-boson and b-quark at the LHC. The constraints on the anomalous couplings 
$C_L$ and $C_R$ have been obtained using the measurements of top-quark decay width and cross-section. Later CP asymmetries were constructed 
for top-pair production process and its decay via the $W$ boson. Using these asymmetries we estimated $2.5\sigma$ level sensitivities to the 
anomalous couplings $C_L$ and $C_R$. These are presented in Tables \ref{cnstrnt_dcyxsec} and \ref{Cons_prodasymm} respectively. The top-quark 
decay width and cross-section set an upper bound on $|C_L|$ of about $(5.9, 2.6) \times 10^{-3}$ respectively at 2.5$\sigma$ C.L. which for 
$|C_R|$ turn out to be $(2.0, 0.4) \times 10^{-3}$ respectively. The corresponding limits on the moduli of the anomalous couplings from the 
production asymmetries with CP-violating phase $\theta=\phi=\frac{\pi}{4}$ are presented in Table~\ref{Cons_prodasymm} for the LHC at a CMS 
energy of $\sqrt{S}$ = 13 TeV with the integrated luminosities of 36.1 fb$^{-1}$ and 139 fb$^{-1}$. The limits are also presented for its 
luminosity intense variant, the HL-LHC, HE-LHC and FCC-hh. Our investigations reveal that upper bounds on $(|C_L|, |C_R|)$ would be of about 
$(1.82,0.03) \times 10^{-4}$ at 2.5$\sigma$ C.L. for the total data collected so far at the LHC with $\sqrt{S}$ = 13 TeV for an 
integrated luminosity of 139 fb$^{-1}$. These for High Luminosity LHC (HL-LHC), High Energy LHC (HE-LHC) and Future circular collider 
(FCC-hh) turned out to be of about $(0.81, 0.006)\times 10^{-4}$, $(0.44, 0.0017)\times 10^{-4}$ and $(0.21, 0.0004)\times 10^{-4}$ for the 
projected luminosities of 3.0 ab$^{-1}$, 12.0 ab$^{-1}$ and 30 ab$^{-1}$ respectively at 2.5$\sigma$ C.L..

\section*{Acknowledgements}
This work was supported in part by University Grant Commission under a Start-Up Grant no. F30-377/2017 (BSR). We thank Surabhi Gupta for 
some valuable discussion. We acknowledge the use of computing facility at the general computing lab of Aligarh Muslim University during the 
initial phase of the work.

\end{document}